\PassOptionsToPackage{quiet}{fontspec}
\PassOptionsToPackage{silence}{Underfull}
\PassOptionsToPackage{silence}{Overfull}
\documentclass[reprint, superscriptaddress, showpacs, amsmath, amssymb, aps, pra,]{revtex4-2}
\usepackage{lipsum}
\usepackage{upgreek}
\usepackage{braket}
\usepackage{silence}
\usepackage{graphicx}
\usepackage{subfigure}
\usepackage{dcolumn}
\usepackage{bm}
\usepackage{hyperref}
\hypersetup{colorlinks=true, citecolor=blue, urlcolor=blue, linkcolor=blue}
\begin{document}
\title{Evaluation of the systematic error induced by quadratic Zeeman effect using hyperfine ground state exchange method in a long-baseline dual-species atom interferometer}

\author{Yu-Hang Ji}
\affiliation{State Key Laboratory of Magnetic Resonance and Atomic and Molecular Physics, Innovation Academy for Precision Measurement Science and Technology, Chinese Academy of Sciences-Wuhan National Laboratory for Optoelectronics, Wuhan 430071, China}

\author{Chuan He}
\affiliation{State Key Laboratory of Magnetic Resonance and Atomic and Molecular Physics, Innovation Academy for Precision Measurement Science and Technology, Chinese Academy of Sciences-Wuhan National Laboratory for Optoelectronics, Wuhan 430071, China}

\author{Si-Tong Yan}
\affiliation{State Key Laboratory of Magnetic Resonance and Atomic and Molecular Physics, Innovation Academy for Precision Measurement Science and Technology, Chinese Academy of Sciences-Wuhan National Laboratory for Optoelectronics, Wuhan 430071, China}

\author{Jun-Jie Jiang}
\affiliation{State Key Laboratory of Magnetic Resonance and Atomic and Molecular Physics, Innovation Academy for Precision Measurement Science and Technology, Chinese Academy of Sciences-Wuhan National Laboratory for Optoelectronics, Wuhan 430071, China}
\affiliation{School of Physical Sciences, University of Chinese Academy of Sciences, Beijing 100049, China}

\author{Jia-Qi Lei}
\affiliation{State Key Laboratory of Magnetic Resonance and Atomic and Molecular Physics, Innovation Academy for Precision Measurement Science and Technology, Chinese Academy of Sciences-Wuhan National Laboratory for Optoelectronics, Wuhan 430071, China}
\affiliation{School of Physical Sciences, University of Chinese Academy of Sciences, Beijing 100049, China}

\author{Lu Zhou}
\affiliation{State Key Laboratory of Magnetic Resonance and Atomic and Molecular Physics, Innovation Academy for Precision Measurement Science and Technology, Chinese Academy of Sciences-Wuhan National Laboratory for Optoelectronics, Wuhan 430071, China}
\affiliation{School of Physical Sciences, University of Chinese Academy of Sciences, Beijing 100049, China}

\author{Lin Zhou}
\email{lzhou@apm.ac.cn}
\affiliation{State Key Laboratory of Magnetic Resonance and Atomic and Molecular Physics, Innovation Academy for Precision Measurement Science and Technology, Chinese Academy of Sciences-Wuhan National Laboratory for Optoelectronics, Wuhan 430071, China}
\affiliation{Hefei National Laboratory, Hefei 230088, China}
\affiliation{Wuhan Institute of Quantum Technology, Wuhan 430206,China}

\author{Xi Chen}
\affiliation{State Key Laboratory of Magnetic Resonance and Atomic and Molecular Physics, Innovation Academy for Precision Measurement Science and Technology, Chinese Academy of Sciences-Wuhan National Laboratory for Optoelectronics, Wuhan 430071, China}
\affiliation{Hefei National Laboratory, Hefei 230088, China}
\affiliation{Wuhan Institute of Quantum Technology, Wuhan 430206,China}

\author{Jin Wang}
\email{wangjin@apm.ac.cn}
\affiliation{State Key Laboratory of Magnetic Resonance and Atomic and Molecular Physics, Innovation Academy for Precision Measurement Science and Technology, Chinese Academy of Sciences-Wuhan National Laboratory for Optoelectronics, Wuhan 430071, China}
\affiliation{Hefei National Laboratory, Hefei 230088, China}
\affiliation{Wuhan Institute of Quantum Technology, Wuhan 430206,China}

\author{Ming-Sheng Zhan}
\email{mszhan@apm.ac.cn}
\affiliation{State Key Laboratory of Magnetic Resonance and Atomic and Molecular Physics, Innovation Academy for Precision Measurement Science and Technology, Chinese Academy of Sciences-Wuhan National Laboratory for Optoelectronics, Wuhan 430071, China}
\affiliation{Hefei National Laboratory, Hefei 230088, China}
\affiliation{Wuhan Institute of Quantum Technology, Wuhan 430206,China}

\email{Second.Author@institution.edu}

\homepage{http://www.Second.institution.edu/~Charlie.Author}

\date{\today}

\begin{abstract}
The systematic error induced by the quadratic Zeeman effect is non-negligible in atom interferometers and must be precisely evaluated. We theoretically analyze the phase shift induced by the Zeeman effect, and use a hyperfine ground state exchange (HGSE) method to evaluate the systematic error in the long-baseline $^{85}$Rb-$^{87}$Rb dual-species atom interferometer due to the quadratic Zeeman effect. Compared to the two evaluation methods, mapping the absolute magnetic field in the interference region and performing phase measurements at different bias fields, the HGSE method could obtain the systematic error in real time in case of slow drifts of either the ambient magnetic field or other systematic effects irrelevant to the hyperfine ground states. To validate the effectiveness of the HGSE method, we also employ the mapping magnetic field method and modulating bias field method independently to cross-check and yield consistent results of three methods within an accuracy of $10^{-11}$ level. The HGSE method is helpful in evaluating and suppressing the quadratic Zeeman-effect-induced systematic error in long-baseline atom interferometer-based precision measurements, such as equivalence principle tests.

\end{abstract}

\maketitle

\section{Introduction}

Atom interferometry has demonstrated remarkable prospects in precision measurements through its developments in three decades, such as atomic gravimeters \cite{peters_high-precision_2001, poli_precision_2011, huang_accuracy_2019, zhong_quantum_2022}, gravity gradiometers \cite{snadden_measurement_1998, sorrentino_sensitive_2010, duan_operating_2014, lyu_compact_2022}, and gyroscopes \cite{gustavson_precision_1997, canuel_six-axis_2006, yao_continuous_2016}, as well as the measurement of fine structure constant \cite{bouchendira_new_2011, parker_measurement_2018} and gravitational constant \cite{fixler_atom_2007, rosi_precision_2014}, and the test of the equivalence principle \cite{fray_atomic_2004, bonnin_simultaneous_2013, schlippert_quantum_2014, tarallo_test_2014, zhou_test_2015, duan_test_2016, barrett_dual_2016, rosi_quantum_2017, zhang_testing_2020, albers_quantum_2020, asenbaum_atom-interferometric_2020, zhou_joint_2021, barrett_testing_2022}. To obtain highly accurate measurement results, systematic errors induced by various effects need to be precisely evaluated. Magnetic effect is one of the important systematic errors in atom interferometers (AIs) \cite{lyu_compact_2022, schlippert_quantum_2014, duan_test_2016, barrett_dual_2016, rosi_quantum_2017, zhang_testing_2020, albers_quantum_2020, asenbaum_atom-interferometric_2020, zhou_joint_2021, barrett_testing_2022}. Commonly, the systematic error induced by the first-order Zeeman effect is avoided by selecting the magnetically insensitive $m_{F}=0$ sublevel of the atoms during the interference process. However, the quadratic Zeeman-effect-induced systematic error is non-negligible.

One evaluation method of the quadratic Zeeman effect is by mapping the absolute magnetic field in the interference region \cite{hu_mapping_2017, deng_precisely_2021}. The mapping magnetic field method is common and robust but limited by the measurement accuracy and spatial resolution of the magnetic field \cite{hu_mapping_2017, deng_precisely_2021, zhou_precisely_2010}. Methods to map the magnetic field mainly include Raman spectroscopy \cite{hu_mapping_2017} and Ramsey interferometer \cite{deng_precisely_2021}. Another evaluation method is by performing phase measurements by the AI at different bias fields and extrapolating to the experimental value \cite{zhang_testing_2020, zhou_joint_2021}. The modulating bias field method does not require mapping the magnetic field, and it gives the systematic error by obtaining the phase shift as a function of the solenoid current, provided that the modulated current does not significantly change the bias field distribution and other systematic effects. Furthermore, measuring the magnetic field gradient in the interference region \cite{asenbaum_atom-interferometric_2020} could be used to evaluate the systematic error induced by the quadratic Zeeman effect. However, it works under the condition that the magnetic field is linearly distributed.

Long-baseline AIs \cite{dimopoulos_testing_2007, zhou_development_2011, hartwig_testing_2015} dramatically increase the accuracy of measurement \cite{asenbaum_atom-interferometric_2020} and extend the range of applications \cite{abend_terrestrial_2023}. However, the evaluation of the systematic error induced by the magnetic field effect encounters significant challenges. On one hand, the magnetic shield of the long-baseline AI \cite{dickerson_high-performance_2012, wodey_scalable_2020, ji_actively_2021} has worse performance than a short baseline one \cite{burt_optimal_2002, milke_atom_2014}. Due to the large length-to-diameter ratio, the axial shielding factor of a 10-m magnetic shield is only about 10 \cite{dickerson_high-performance_2012, ji_actively_2021}. This makes long-baseline AIs more susceptible to vertical ambient magnetic fields. The magnetic field environment is complex due to the presence of metros, elevators, vehicles, and instruments near a laboratory, and the vertical variation even reaches tens of milligausses. On the other hand, long-baseline AIs require the systematic error induced by magnetic field effect to be evaluated with a higher accuracy \cite{asenbaum_atom-interferometric_2020, zhou_joint_2021}. The 10-m AI takes over ten times longer than that with compact devices to map the absolute magnetic field in the vacuum using atoms, which means a more extended period for evaluating the systematic error. The evaluation methods mentioned above all require a stable magnetic field. Considering the shielding performance of the large length-to-diameter magnetic shield, the effects of metros and other factors could severely limit the measurement precision of the AI and the evaluation accuracy of the Zeeman-effect-induced systematic error, or make high-precision measurements possible for only a small portion of the day. Therefore, evaluating the systematic error caused by the magnetic field effect in real time is essential for long-baseline AIs.

In this paper, we realize an evaluation method applicable to the long-baseline dual-species AI based on our previous work \cite{zhou_joint_2021}. The method does not require precise measurement of the magnetic field, and it could evaluate the systematic error caused by the Zeeman effect in real time, even if the ambient magnetic field and other systematic effects irrelevant to the hyperfine ground states change slowly. First, we derive an analytical expression for the phase shift induced by the Zeeman effect. Then, we realize a method called hyperfine ground state exchange (HGSE) by alternating the hyperfine ground state between two consecutive shots. Finally, considering the effectiveness of this method, we employ three independent evaluation methods to cross-check the quadratic Zeeman effect. In addition, we demonstrate that the HGSE method could evaluate the quadratic Zeeman-effect-induced systematic error in real time, in case of slow drifts of the ambient magnetic field and other systematic effects irrelevant to the hyperfine ground states. The paper is organized as follows. In Sec.\,\ref{section2}, we analyze the phase shift induced by the quadratic Zeeman effect. The experimental setup and procedure are briefly introduced in Sec.\,\ref{section3}. In Sec.\,\ref{section4}, we demonstrate the evaluations of the quadratic Zeeman-effect-induced systematic error. In Sec.\,\ref{section5}, we discusses the robustness of the HGSE method. Section\,\ref{section6} summarizes our main results and provides an outlook.

\section{THEORETICAL ANALYSIS}\label{section2}

\subsection{Zeeman shift}

The Zeeman shifts of the energy levels of alkali atoms, including all magnetic substates of the two ground hyperfine levels, are precisely described by the Breit-Rabi formula \cite{breit_measurement_1931}. According to this formula, the energy shift corresponding to the ground-state magnetic sublevel $\ket{F, m_{F}}$ due to the small magnetic field $B$ can be written as
\begin{align}
    \Delta E_{B} ={}& (g_{I}\pm \frac{g_{J}-g_{I}}{2I+1})\mu _{B} m_{F} B \notag \\
    &\pm \left( 1-\frac{4m_{F}^{2}}{(2I+1)^{2}} \right) \frac{(g_{J}-g_{I})^{2}\mu _{B}^{2}}{4\Delta E_{\mathrm{hfs}}}B^{2},
    \label{eq1}
\end{align}
where $g_{J}$ and $g_{I}$ are the electronic and nuclear Landé $g$ factors, $I$ is the total nuclear angular momentum, which is $5/2$ for $^{85}$Rb and $3/2$ for $^{87}$Rb, $\mu _{B}$ is the Bohr magneton, $m_{F}=0, \pm 1, \cdots\pm F$ are the projections of total angular momentum on the quantization axis, $\Delta E_{\mathrm{hfs}} = A_{\mathrm{hfs}}(I+1/2)$ is the hyperfine splitting. Note that the Zeeman shift is positive for atoms in the upper ground state (UGS) and negative for atoms in the lower ground state (LGS) \cite{zhou_joint_2021}. Therefore, the hyperfine sublevel $\ket{F, m_{\mathrm{F}}=0}$ has no first-order Zeeman shift, the corresponding frequency shift is defined as
\begin{equation}
    \Delta \omega _{i\text{-}F} = \pm \frac{(g_{J}-g_{I})^{2}\mu _{B}^{2}}{4\hbar \Delta E_{\mathrm{hfs}}}B^{2} = 2\pi \alpha _{i\text{-}F}B^{2},
    \label{eq2}
\end{equation}
where $\hbar$ is the reduced Planck constant, $\alpha _{i\text{-}F}$ is the quadratic Zeeman coefficient of isotope $i$ with hyperfine ground state $F$. For the $\ket{F=2, m_{F}=0}$ and $\ket{F=3, m_{F}=0}$ hyperfine levels of the $5^{2}\text{S}_{1/2}$ ground state of $^{85}$Rb \cite{steck_rubidium85_2023}, the calculated coefficients are $\alpha _{85\text{-}2} = -646.99\,\mathrm{Hz/G^{2}}$ and $\alpha _{85\text{-}3} = 646.99\,\mathrm{Hz/G^{2}}$, respectively. For the $\ket{F=1, m_{\mathrm{F}}=0}$ and $\ket{F=2, m_{\mathrm{F}}=0}$ hyperfine levels of the $5^{2}\text{S}_{1/2}$ ground state of $^{87}$Rb \cite{steck_rubidium87_2023}, the calculated coefficient $\alpha _{87\text{-}1} = -287.57\,\mathrm{Hz/G^{2}}$ and $\alpha _{87\text{-}2} = 287.57\,\mathrm{Hz/G^{2}}$, respectively.

\subsection{Quadratic Zeeman-effect-induced phase shift}

While the atoms in the $\ket{F, m_{F}=0}$ sublevel are in free fall in a magnetically shielded region \cite{zhou_test_2015,zhou_joint_2021}, we perform a Raman Mach-Zehnder interferometer with the $\pi/2\text{-}\pi\text{-}\pi/2$ Doppler-sensitive configuration, and the interferometer duration is $2T$. When the interference process is finished, the phase shift $\Delta \phi$ caused by the quadratic Zeeman effect can be written as
\begin{equation}
    \Delta \phi = 2\pi \textstyle\int_{0}^{2T} \lbrace \alpha _{i\text{-}F}^{(u)}B^{2}[z^{(u)}(t)]-\alpha _{i\text{-}F}^{(d)}B^{2}[z^{(d)}(t)] \rbrace dt,
    \label{eq3}
\end{equation}
where $B[z^{(u)}(t)]$ and $B[z^{(d)}(t)]$ are magnetic fields at the position $z^{(u)}(t)$ of the upward path and at the position $z^{(d)}(t)$ of the downward path, respectively, $\alpha _{i\text{-}F}^{(u)}$ and $\alpha _{i\text{-}F}^{(d)}$ are the quadratic Zeeman coefficient in the upward path and downward path, respectively.

In the single Raman diffraction (SRD) scheme, the Raman pulse can change the momentum of the atom and, simultaneously, its hyperfine ground state, so $\alpha _{i\text{-}F}^{(u)} = -\alpha _{i\text{-}F}^{(d)}$. However, the atoms stay at the same hyperfine ground state in the double Raman diffractiom (DRD) scheme, so $\alpha_{i\text{-}F}^{(u)}=\alpha_{i\text{-}F}^{(d)}$. This interferometric method, thanks to the identical quadratic Zeeman coefficient during the interference process, is more insensitive to the inhomogeneous magnetic field \cite{berg_composite-light-pulse_2015, olivares-renteria_quantum_2020}. The quadratic Zeeman-effect-induced phase shift in the DRD scheme is
\begin{equation}
    \Delta \phi _{i\text{-}F} = 2\pi \alpha _{i\text{-}F}\textstyle\int_{0}^{2T} \lbrace B^{2}[z^{(u)}(t)]-B^{2}[z^{(d)}(t)] \rbrace dt.
    \label{eq4}
\end{equation}
The phase shift $\Delta \phi _{i\text{-}F}$ is proportional to $\alpha _{i\text{-}F}$ and mainly arises from the magnetic field inhomogeneity between the upward and downward paths of the interferometer.

In the SRD scheme, to decrease the influence of the quadratic Zeeman effect according to Eq.\,(\ref{eq3}), the atomic trajectory during the first half and second half of the interferometer path should be almost symmetric. Whereas, for a typical fountain configuration, we have to make the interferometer slightly asymmetric so as to prevent the central Raman pulse from driving unwanted Doppler-insensitive transitions \cite{peters_high-precision_2001, hu_mapping_2017}. However, from Eq.\,(\ref{eq4}), the quadratic Zeeman phase shift in the DRD scheme is insensitive to the symmetry of interferometer paths, and the magnetic field inhomogeneity between the upward and downward paths is usually smaller than the inhomogeneity between the first-half and second-half paths. Therefore, the quadratic Zeeman shift of the DRD AI is generally much smaller than that of the SRD AI.

\section{EXPERIMENT}\label{section3}

\subsection{Experimental apparatus}

\begin{figure}[b]
  \centering
  \includegraphics[width=0.45\textwidth]{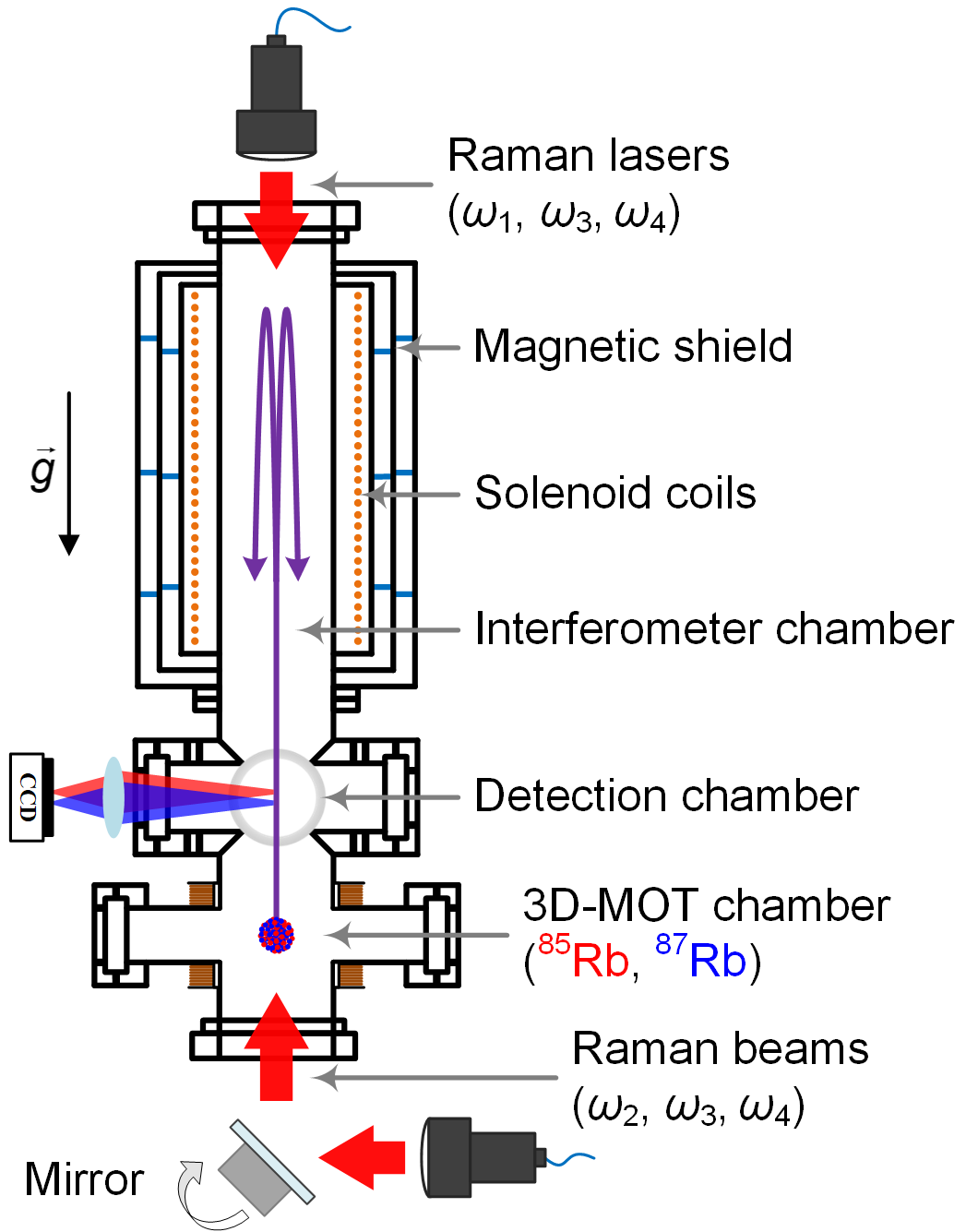}
  \caption{Schematic diagram of the experimental setup.}
  \label{fig1}
\end{figure}

The schematic diagram of the experimental setup shown in Fig.\,\ref{fig1} is similar to our previous system \cite{zhou_joint_2021}, except that the detection scheme of phase shear readout \cite{zhou_toward_2022, yan_absolute-phase-shift_2023}, where a CCD camera images the atomic density distribution with vertical fluorescence beams. A group of Raman lasers ($\omega _{1}$, $\omega _{3}$, and $\omega _{4}$), blow-away, and repumping beams propagate downward through the top window of the vacuum chamber. Another group of Raman beams ($\omega _{2}$, $\omega _{3}$, and $\omega _{4}$) propagate upward through the bottom window of the vacuum chamber. These four Raman lasers compose the four-wave double-diffraction Raman transition (4WDR) scheme. An 11.4 m-long magnetic shielding system \cite{ji_actively_2021} is achieved by a combination of passive shielding using three-layer cylindrical permalloy and active compensation with external, internal, and solenoid coils. The shield provides an axial shielding factor of less than 10 and a transverse shielding factor of more than $1\times10^{4}$, due to the large length-to-diameter ratio. The bias magnetic field is supplied by a solenoid inside the magnetic shielding system, which defines the quantization axis.

Here we give a brief introduction to the experimental process. Firstly, $^{85}$Rb and $^{87}$Rb atoms are cooled and trapped by the three-dimensional magneto-optical trap (3D-MOT). Secondly, the cold $^{85}$Rb and $^{87}$Rb atoms are launched simultaneously by a moving molasses process to form atom fountains. Thirdly, after entering the magnetic shielding zone, the atoms are prepared to the magnetically insensitive state ($m_{F}=0$) and selected with a narrow vertical velocity distribution, which is achieved by the Doppler-sensitive Raman beams propagated along the quantization axis.
Afterward, a $\pi/2$-$\pi$-$\pi/2$ Raman pulse sequence is applied to split, reflect, and recombine the atomic wave packet, which is separated by a free evolution time of $T$. Finally, we get the differential phase between $^{85}$Rb and $^{87}$Rb by the phase shear readout with the internal state labeling detection.

\subsection{Raman-type atom interferometer}

\begin{figure}[b]
  \centering
  \includegraphics[width=0.45\textwidth]{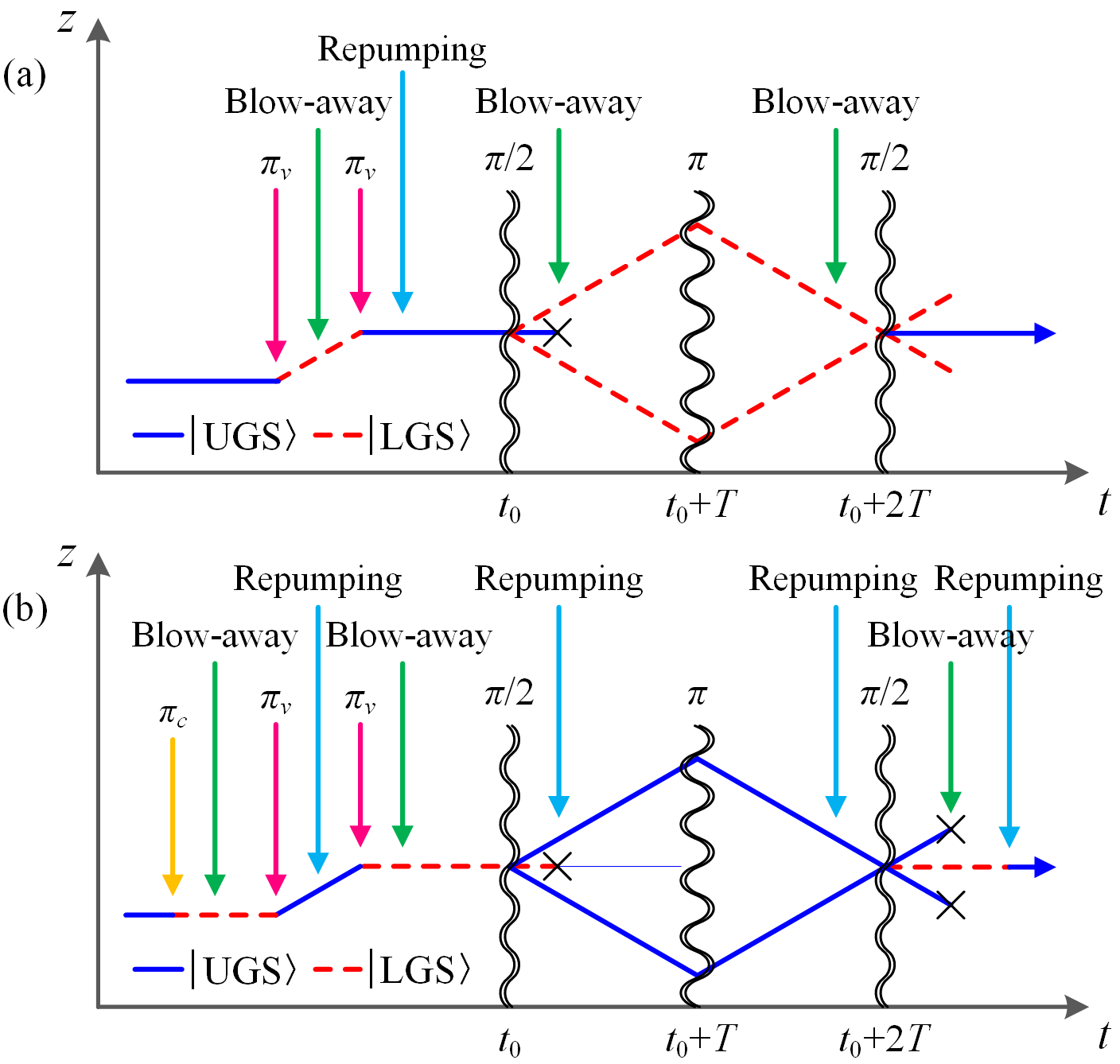}
  \caption{Space-time diagram of the double-diffraction Raman-type atom interferometer: (a) the lower ground state atom interferometer, (b) the upper ground state atom interferometer.}
  \label{fig2}
\end{figure}

The Raman-type atom interferometer is based on the stimulated Raman transition \cite{peters_high-precision_2001, kasevich_atomic_1991}. In brief, the SRD scheme realizes a configuration with asymmetric momentum-space splitting of $\hbar k_{\mathrm{eff}}$ and two hyperfine ground states \cite{berg_composite-light-pulse_2015, hartmann_regimes_2020}. In the DRD scheme \cite{zhou_test_2015, hartmann_regimes_2020, leveque_enhancing_2009, malossi_double_2010}, the atom interacts with two laser pairs and consequently diffraction in both directions to achieve a symmetric momentum-space splitting of $2\hbar k_{\mathrm{eff}}$, in which the atomic wave packets are in the same hyperfine ground state. Compared to the SRD, the resonance condition does not change, except for the Rabi oscillations with an effective Rabi frequency of $\sqrt{2}\Omega _{\mathrm{eff}}$, where $\Omega _{\mathrm{eff}}$ is the effective two-photon Rabi frequency for SRD \cite{leveque_enhancing_2009}. As shown in Fig.\,\ref{fig2}, the first $\pi /2$ pulse with duration $\tau _{\pi /2}^{(D)} = \pi / (\sqrt{2}\Omega _{\mathrm{eff}})$ excites the initial state $\ket{F=a, \boldsymbol{p}=\boldsymbol{p}_{\mathrm{0}}}$ to two states $\ket{F=b, \boldsymbol{p}=\boldsymbol{p}_{\mathrm{0}}+\hbar \boldsymbol{k}_{\mathrm{eff}}}$ and $\ket{F=b, \boldsymbol{p}=\boldsymbol{p}_{\mathrm{0}}-\hbar \boldsymbol{k}_{\mathrm{eff}}}$, where $\boldsymbol{k}_{\mathrm{eff}}=\boldsymbol{k}_{\mathrm{1}}-\boldsymbol{k}_{\mathrm{2}}$ is the effective wave vector. The $\pi$ pulse with duration $\tau _{\pi}^{(D)} = \sqrt{2}\pi / \Omega _{\mathrm{eff}}$ acts as a mirror in each path to reflect the states with $\ket{F=b, \boldsymbol{p}=\boldsymbol{p}_{\mathrm{0}}+\hbar \boldsymbol{k}_{\mathrm{eff}}}$ $\rightarrow$ $\ket{F=b, \boldsymbol{p}=\boldsymbol{p}_{\mathrm{0}}-\hbar \boldsymbol{k}_{\mathrm{eff}}}$ and $\ket{F=b, \boldsymbol{p}=\boldsymbol{p}_{\mathrm{0}}-\hbar \boldsymbol{k}_{\mathrm{eff}}}$ $\rightarrow$ $\ket{F=b, \boldsymbol{p}=\boldsymbol{p}_{\mathrm{0}}+\hbar \boldsymbol{k}_{\mathrm{eff}}}$. Finally, atomic wave packets are recombined due to the second $\pi/2$ pulse with duration $\tau _{\pi /2}^{(D)}$.

The LGS and UGS AIs using $^{85}$Rb and $^{87}$Rb atoms compose four combination pairs for differential measurements. In this paper, we implement two types of interferometers with $^{85}$Rb and $^{87}$Rb staying either in the LGS or the UGS, calling LGS-AI and UGS-AI, to form the HGSE method. In the LGS-AI, atoms stay at states $^{85}$Rb $\ket{F=2}$ and $^{87}$Rb $\ket{F=1}$ during the interference process as shown in Fig.\,\ref{fig2}\textcolor{blue}{(a)}. A $\pi_{v}$-blow-away-$\pi_{v}$-repumping pulse sequence is applied to $^{85}$Rb and $^{87}$Rb for state preparation and velocity selection, then atoms are in states $^{85}$Rb $\ket{F=3}$ and $^{87}$Rb $\ket{F=2}$. Here, the $\pi_{v}$ pulse is a Doppler-sensitive single-diffraction Raman pulse used to transfer narrow-velocity atoms. We use two $\pi_{v}$ pulses to make the atom's velocity the same as the fountain to reduce the influence of the pulses on the atom's velocity and trajectory. The blow-away pulse is used to remove the unwanted atoms residing in states $^{85}$Rb $\ket{F=3}$ and $^{87}$Rb $\ket{F=2}$, and a repumping pulse is applied to repump atoms from $^{85}$Rb $\ket{F=2}$ and $^{87}$Rb $\ket{F=1}$ to $^{85}$Rb $\ket{F=3}$ and $^{87}$Rb $\ket{F=2}$. A $\pi/2$-blow-away-$\pi$-blow-away-$\pi/2$ pulse sequence is applied to realize $^{85}$Rb $\ket{F=2}$-$^{87}$Rb $\ket{F=1}$ dual-species LGS-AI. Here, the blow-away pulses are used to remove the unwanted atoms in $^{85}$Rb $\ket{F=3}$ and $^{87}$Rb $\ket{F=2}$ states in the middle path.

In the UGS-AI, atoms stay at states $^{85}$Rb $\ket{F=3}$ and $^{87}$Rb $\ket{F=2}$ during the interference process as shown in Fig.\,\ref{fig2}\textcolor{blue}{(b)}. A $\pi_{c}$-blow-away-$\pi_{v}$-repumping-$\pi_{v}$-blow-away pulse sequence is applied to $^{85}$Rb and $^{87}$Rb for state preparation and velocity selection, then atoms are in states $^{85}$Rb $\ket{F=2}$ and $^{87}$Rb $\ket{F=1}$. Here the $\pi_{c}$ pulse, a copropagating Doppler-insensitive Raman pulse, is used to only transfer atoms from $^{85}$Rb $\ket{F=3}$ and $^{87}$Rb $\ket{F=2}$ to $^{85}$Rb $\ket{F=2}$ and $^{87}$Rb $\ket{F=1}$, but nearly does not change the velocities. The purpose of other laser pulses is the same as that described in Fig.\,\ref{fig2}\textcolor{blue}{(a)}. In order to overlap completely with the atomic trajectories of the LGS-AI, the $\pi_{v}$ pulses are at the same moment. A $\pi/2$-repumping-$\pi$-repumping-$\pi/2$ pulse sequence is applied to realize $^{85}$Rb $\ket{F=3}$-$^{87}$Rb $\ket{F=2}$ dual-species UGS-AI. Here, the repumping pulse is used to deviate the unwanted atoms in $^{85}$Rb $\ket{F=2}$ and $^{87}$Rb $\ket{F=1}$ states from the interference loop. The last blow-away-repumping pulse sequence removes the remaining atoms in the $^{85}$Rb $\ket{F=3}$ and $^{87}$Rb $\ket{F=2}$ states and then pumps the atoms in the $^{85}$Rb $\ket{F=2}$ and $^{87}$Rb $\ket{F=1}$ states to $^{85}$Rb $\ket{F=3}$ and $^{87}$Rb $\ket{F=2}$ for detection.

\section{EVALUATIONS OF QUADRATIC ZEEMAN-EFFECT-INDUCED SYSTEMATIC ERROR}\label{section4}

\subsection{Hyperfine ground state exchange method}

In our previous work \cite{zhou_joint_2021}, as shown in Eq.\,(\ref{eq2}), we took advantage of the opposite sign of quadratic Zeeman coefficients of the lower and upper ground states to give systematic errors for four combination pairs of specified mass and internal energy. A similar consideration was realized by Panda \textit{et al}. \cite{panda_atomic_2023}, who suppressed the effect of strong environmental magnetic fields and field gradients using atoms in the two hyperfine states as co-magnetometers. Due to the quadratic Zeeman effect, the inhomogeneous magnetic field induces the systematic shift, and the uncertainty of the magnetic field is part of responsible for systematic uncertainty. We will briefly describe the HGSE method below.
 
The differential phase is shifted by systematic effects. These systematic phase shifts can be sorted into two classes of error sources \cite{louchet-chauvet_influence_2011}, either been dependent $(\Delta \phi _{\mathrm{dep}})$ or independent $(\Delta \phi _{\mathrm{indep}})$ on the hyperfine ground state $F$. The differential phase of the dual-species AI can thus be expressed as $\Delta \phi _{F} = \Delta \phi _{g} + \Delta \phi _{\mathrm{dep}} + \Delta \phi _{\mathrm{indep}}$, where $\Delta \phi _{g}= \Delta k_{\mathrm{eff}}gT^{2} + k_{\mathrm{eff}}\Delta gT^{2}$, and the first term is caused by the difference of effective wave vectors $k_{\mathrm{eff}}$ of the atoms, the second term is used to test the weak equivalence principle, which is caused by the potential relative acceleration $\Delta g$ between specified mass and internal energy of atoms. Taking that into account, the measurement procedure we use interleaved differential phase measurements with the LGS and UGS:
\begin{equation}
\begin{split}
    & \Delta \phi _{\mathrm{LGS}} = \Delta \phi _{g} - \Delta \phi _{\mathrm{dep}} + \Delta \phi _{\mathrm{indep}} \\
     & \Delta \phi _{\mathrm{UGS}} = \Delta \phi _{g} + \Delta \phi _{\mathrm{dep}} + \Delta \phi _{\mathrm{indep}}.
     \label{eq5}
\end{split}
\end{equation}

Half-difference and half-sum of successive $\Delta \phi _{\mathrm{LGS}}$ and $\Delta \phi _{\mathrm{UGS}}$ measurements allow us to separate $\Delta \phi _{g} + \Delta \phi _{\mathrm{indep}}$. $\Delta \phi _{\mathrm{indep}}$ originates from effects related to perturbations of the external degrees of freedom of the atoms (such as gravity gradient, Coriolis effect, and wave-front aberration) and from the Raman laser phase shifts. $\Delta \phi _{\mathrm{dep}}$ mainly includes the quadratic Zeeman shift and ac Stark shift. Here, the ac Stark shift is caused by Raman, blow-away, and repumping lasers. To cancel the total ac Stark shift, the magic intensity ratio of the four Raman lasers in dual-species Raman transitions is controlled as $I _{1} : I _{2}: I _{3}: I _{4} = 1.00 : 1.00 : 3.05 : 14.3$. Testing the equivalence principle at $10^{-11}$ level in our experiment, the influence of the ac Stark shift could be neglected in the $\Delta \phi _{\mathrm{dep}}$. The LGS-AI and UGS-AI are alternated at the two consecutive shots using the HGSE method. Therefore, the quadratic Zeeman phase shift ($\Delta \phi _{\mathrm{Zeeman}} \approx \Delta \phi _{\mathrm{dep}}$) in the LGS-AI is obtained by
\begin{equation}
    \Delta \phi _{\mathrm{Zeeman}} = (\Delta \phi _{\mathrm{LGS}} - \Delta \phi _{\mathrm{UGS}})/2.
    \label{eq6}
\end{equation}

In our experiment, we implement a simultaneous $^{85}$Rb-$^{87}$Rb dual-species AI to test the equivalence principle, the phase noises and vibrational noise are suppressed by the 4WDR (DRD) method \cite{zhou_test_2015, zhou_joint_2021}. The typical experimental parameters for this experiment are as follows, the launch velocity is $v_{0} = 7.8\,\mathrm{m/s}$, the $\pi$ pulse duration is $\tau _{\pi}^{(S)} = \pi / \Omega _{\mathrm{eff}} = 60\,\mathrm{\mu s}$ for the SRD scheme, and $\tau _{\pi}^{(D)} = \sqrt{2}\pi / \Omega _{\mathrm{eff}} = 84\,\mathrm{\mu s}$ for the DRD scheme, the time of the first $\pi/2$ Raman pulse is $t_{0} = 0.33\,\mathrm{s}$ after launch, the time interval between $\pi/2 \text{-} \pi \text{-} \pi/2$ Raman pulses is $T = 0.45\,\mathrm{s}$. Correspondingly, the height of the fountain apex is $h_{\mathrm{Apex}}\approx 3.12\,\mathrm{m}$, and the heights of three Raman pulses are $h_{\mathrm{\pi/2}}\approx 1.96\,\mathrm{m}$, $h_{\mathrm{\pi}}\approx 3.11\,\mathrm{m}$, and $h_{\mathrm{\pi/2}}^{\prime}\approx 2.29\,\mathrm{m}$, respectively.

\begin{figure}[b]
  \centering
  \includegraphics[width=0.45\textwidth]{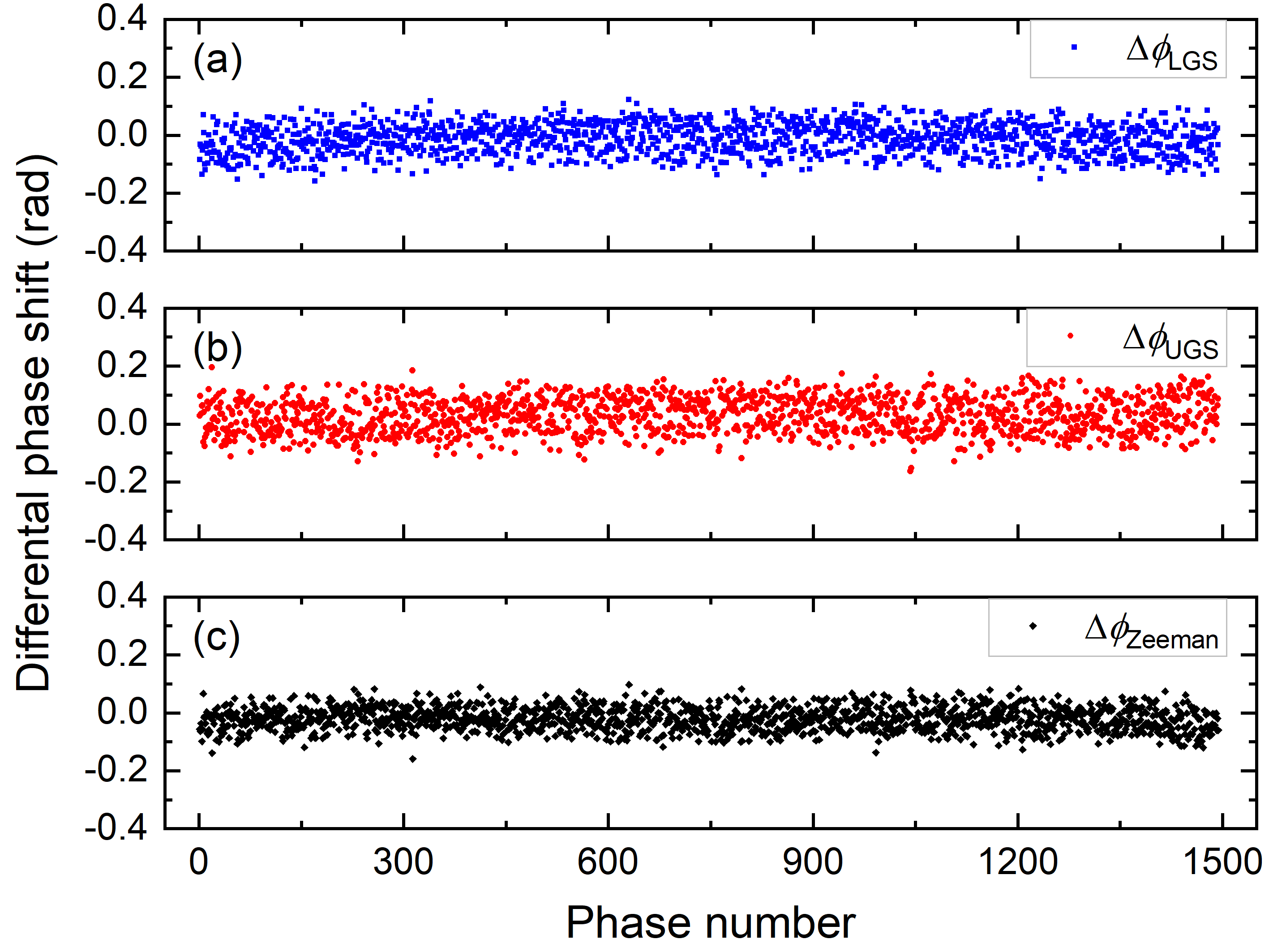}
  \caption{The evaluation results by the HGSE method with the solenoid current of $200\, \mathrm{mA}$. The three sets of data (from top to bottom) are the phases $\Delta \phi_\mathrm{LGS}$, $\Delta \phi_\mathrm{UGS}$, and $(\Delta \phi_\mathrm{LGS}-\Delta \phi_\mathrm{LGS})/2$, respectively. ($\Delta \phi_\mathrm{LGS}$ and $\Delta \phi_\mathrm{UGS}$ are subtracted by $\Delta k_{\mathrm{eff}} g T^{2}$)}
  \label{fig3}
\end{figure}

We apply the HGSE method by alternating the LGS-AI and UGS-AI consecutively. The atomic trajectories of LGS-AI and UGS-AI overlap, achieved by a preparative pulse sequence controlling the input atoms precisely, as shown in Fig.\,\ref{fig2}. The quadratic Zeeman-effect-induced systematic error of the LGS-AI is evaluated by the HGSE method with the typical parameters and a certain solenoid current. All of the phases $\Delta \phi_{\mathrm{LGS}}$ and $\Delta \phi_{\mathrm{UGS}}$ discussed in this paper are subtracted by $\Delta k_{\mathrm{eff}} g T^{2}$, as shown in Fig.\,\ref{fig3}. We evaluate the systematic error of $(-13.0 \pm 4.0) \times 10^{-11}$ at $200\,\mathrm{mA}$ with 1500 measurements. The shift provided here is based on the assumption that other terms dependent on the hyperfine ground state exhibit negligible phase shifts, including the ac Stark shift and the possible violations of the equivalence principle related to internal energy. These can be further distinguished through experiments, such as by precisely modulating the bias magnetic field, laser intensity, and state combination pairs of atoms. The uncertainty presented here is primarily limited by the resolution of the AI.

\subsection{Mapping magnetic field method and modulating bias field method}

To validate the effectiveness of the HGSE method, we use two other independent methods to cross-check the evaluation results. Firstly, we evaluate the quadratic Zeeman-effect-induced systematic error by mapping the absolute magnetic field in the interference region. To map the absolute magnetic field in vacuum by the Raman spectroscopy method \cite{hu_mapping_2017, zhou_precisely_2010}, which has to irradiate the $^{87}$Rb atoms with Raman pulses at different time on the atom’s trajectory. We use the magic intensity ratio of the Raman beams with a $500\, \mathrm{\mu s}$ pulse length and $400\,\mathrm{Hz}$ frequency step to map the magnetic field inside the 2-m-length interferometer chamber.

\begin{figure*}[t]
  \centering
  \includegraphics[width=0.45\textwidth]{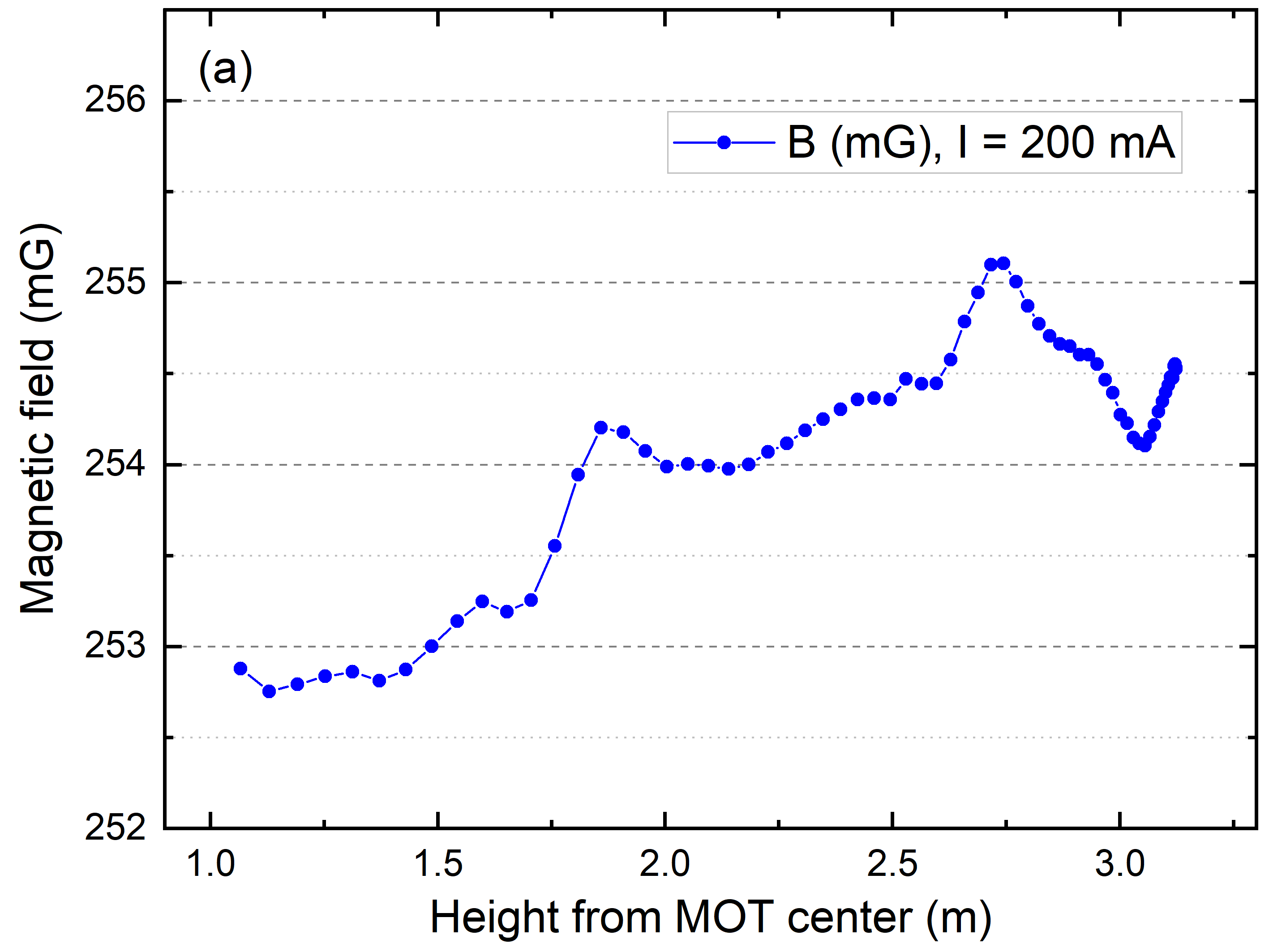}
  \includegraphics[width=0.45\textwidth]{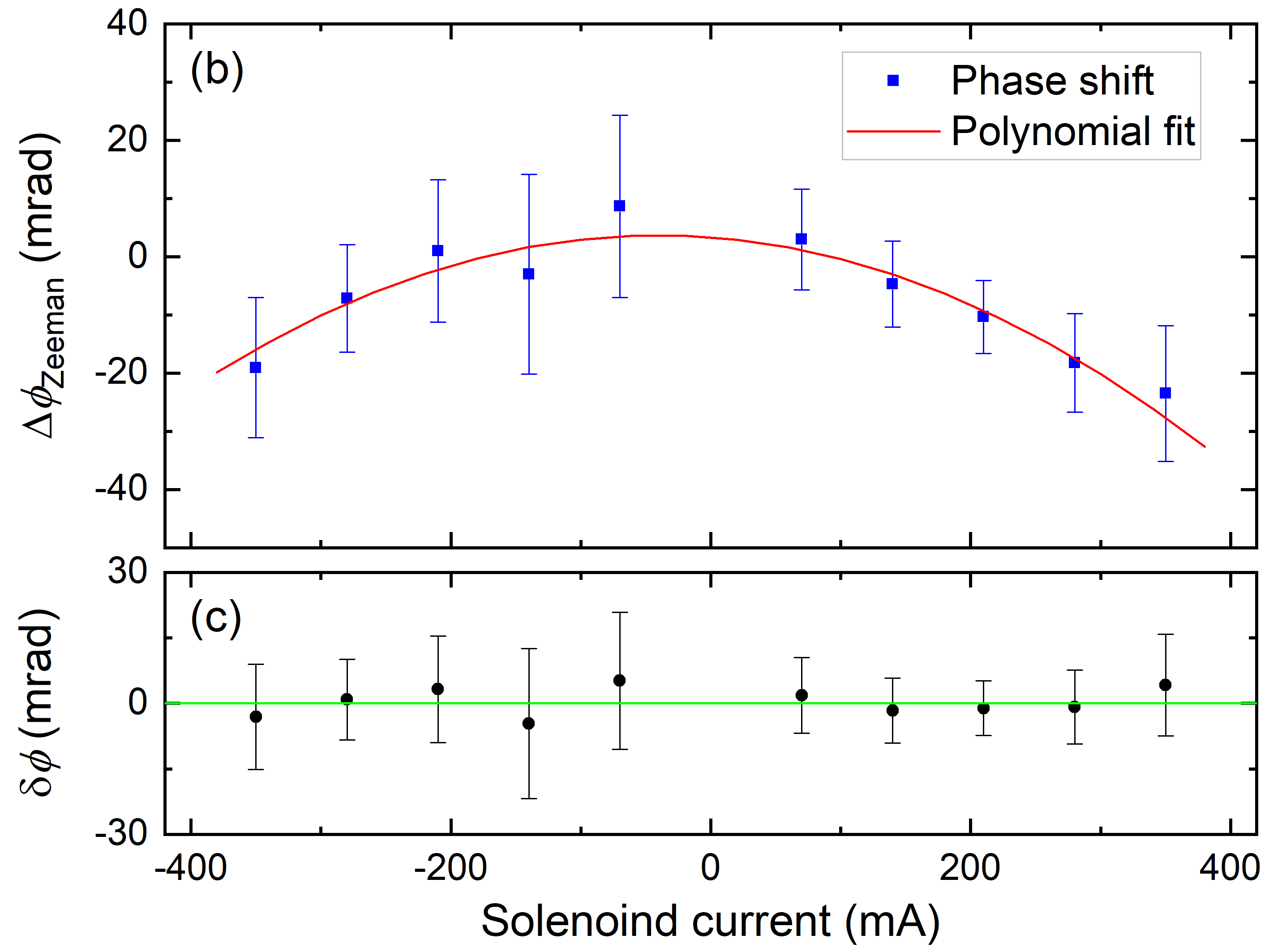}
  \caption{Cross-check with two methods. (a) The bias magnetic field in the interference region with the solenoid current at $200\, \mathrm{mA}$. (b) Evaluation based on phase measurements by modulating the bias magnetic field. The differential phase $\Delta \phi_{\mathrm{Zeeman}}$ (blue squares) is measured by the atom interferometer with different solenoid currents. The red line is the quadratic polynomial fit result. (c) The black dots represent the $\delta \phi$ after correcting the quadratic Zeeman effect.}
  \label{fig4}
\end{figure*}

As shown in Fig.\,\ref{fig4}\textcolor{blue}{(a)}, the result is a measured magnetic field with the solenoid current set at $200\,\mathrm{mA}$ at night, where the height is referred to the center of the MOT chamber. The mean value is $254.1\,\mathrm{mG}$, the inhomogeneity is mainly caused by the compensation coils and defective joints in the solenoid coil. The measurement uncertainty is less than $7.0\, \mathrm{\mu G}$, which mainly originates from detection noise and fitting errors. The solenoid magnetic field is driven by a laser diode current driver (Thorlabs LDC205C) with a current drift of $\sim\!10\,\mathrm{\mu A}$. From Eq.\,(\ref{eq4}), we can infer that the quadratic Zeeman-effect-induced systematic error of the LGS-AI is $(-12.6\pm0.9)\times10^{-11}$ at $200\,\mathrm{mA}$ by the interpolation integral of the magnetic field map. The uncertainty is inferred by the variation of the ambient magnetic field mainly caused by the metros, which we describe in more detail in Sec.\,\ref{section5}. If there were greater variations in the magnetic field and gradient, the uncertainty would be higher.

In addition, we evaluate the quadratic Zeeman-effect-induced systematic error by performing phase measurements in the LGS-AI at different bias fields \cite{zhang_testing_2020, zhou_joint_2021}. The bias field consists of the residual magnetic field inside the shield and the solenoid magnetic field, so it is linearly dependent on the solenoid current. Here, we implement the interferometer with the typical parameters and the solenoid current from $-350$ to $350\,\mathrm{mA}$ spacing $70\,\mathrm{mA}$ (except for $0\,\mathrm{mA}$, in which the transition peaks of Raman spectroscopy are inseparable), respectively, and repeat 250 times with each current. A negative value of current indicates that its direction is opposite to the positive value. From Eq.\,(\ref{eq4}), there is a quadratic function relationship between the differential phase shift and the bias field. Therefore, as shown in Fig.\,\ref{fig4}\textcolor{blue}{(b)}, the differential phase shift has a quadratic function relationship with the solenoid current. Figure\,\ref{fig4}\textcolor{blue}{(c)} shows the residual shift (black dots) by subtracting the quadratic polynomial fit. The phase deviates from the quadratic curve due to measurement errors and the absence of a perfect linear correlation between the magnetic field and the solenoid current. Extrapolating the solenoid current to $200\, \mathrm{mA}$, the quadratic Zeeman-effect-induced systematic error of the LGS-AI is $(-12.9\pm1.1)\times10^{-11}$, where the uncertainty is the standard deviation of the weighted mean of these measurements. Here, the systematic effect is amplified by modulating the solenoid current, so the uncertainty is beyond the interferometer resolution. When the accuracy is further improved, the nonlinear correlation between the magnetic field distribution and the solenoid current would be a challenge to evaluate.

\subsection{Comparison of three evaluation methods}

The systematic shifts of the LGS-AI obtained from the three evaluation methods (the HGSE method, the mapping magnetic field method, and the modulating bias field method) are $(-13.0\sim-12.6)\times10^{-11}$, within the uncertainty range. These three evaluation methods are carried out independently, and the cross-check results are consistent, which indicates that these methods are accurate with the current experimental precision and parameters.

\section{DISCUSSION}\label{section5}

\begin{figure}[b]
  \centering
  \includegraphics[width=0.45\textwidth]{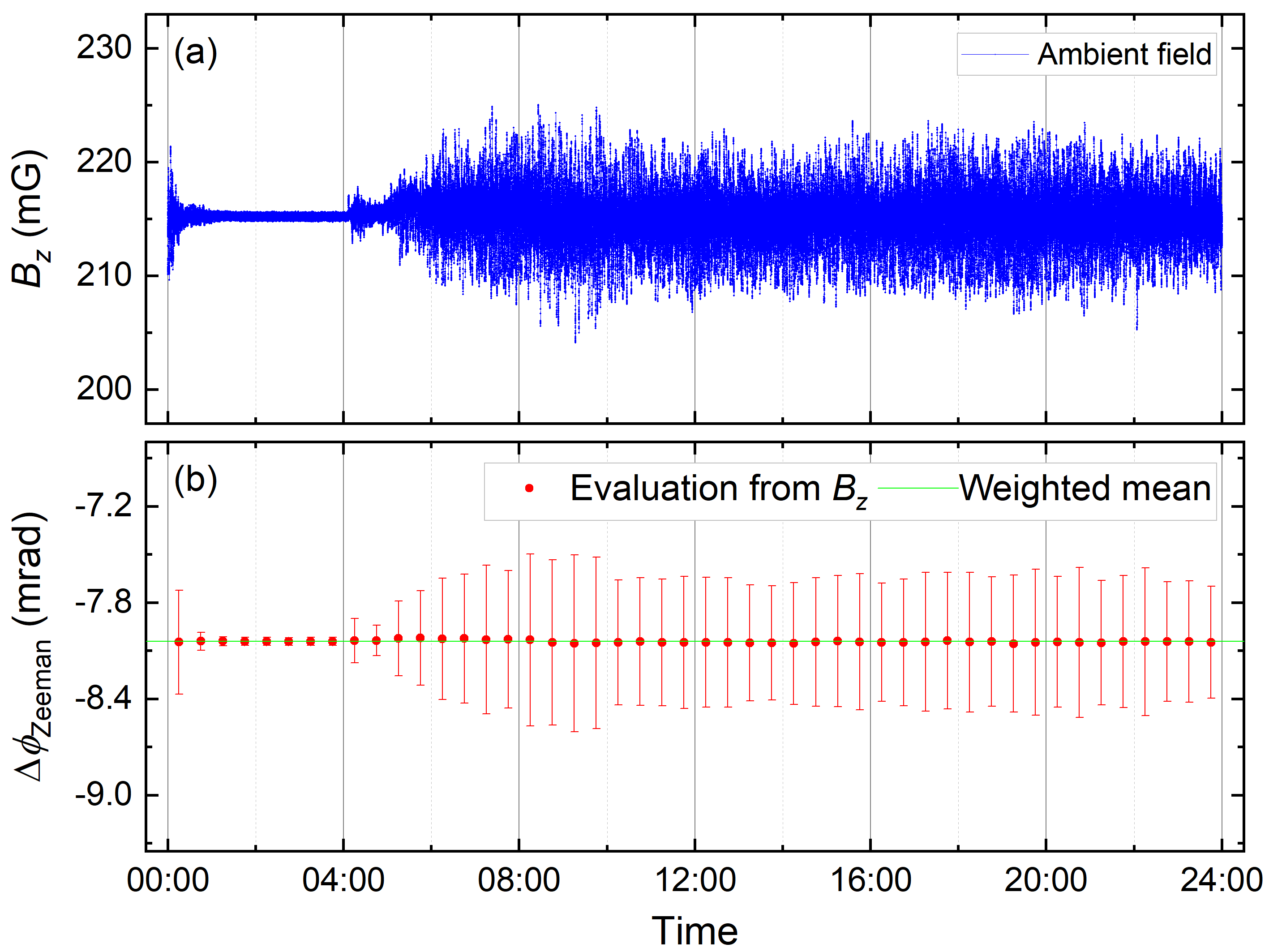}
  \caption{The ambient magnetic field and the quadratic Zeeman phase shift. (a) The measured ambient magnetic field in the vertical direction for $24\,\mathrm{hours}$. (b) The corresponding inferred quadratic Zeeman phase shifts. Each point corresponds to the estimated residual magnetic field over $30\,\mathrm{minutes}$.}
  \label{fig5}
\end{figure}

\begin{figure*}[t]
  \centering
  \includegraphics[width=0.45\textwidth]{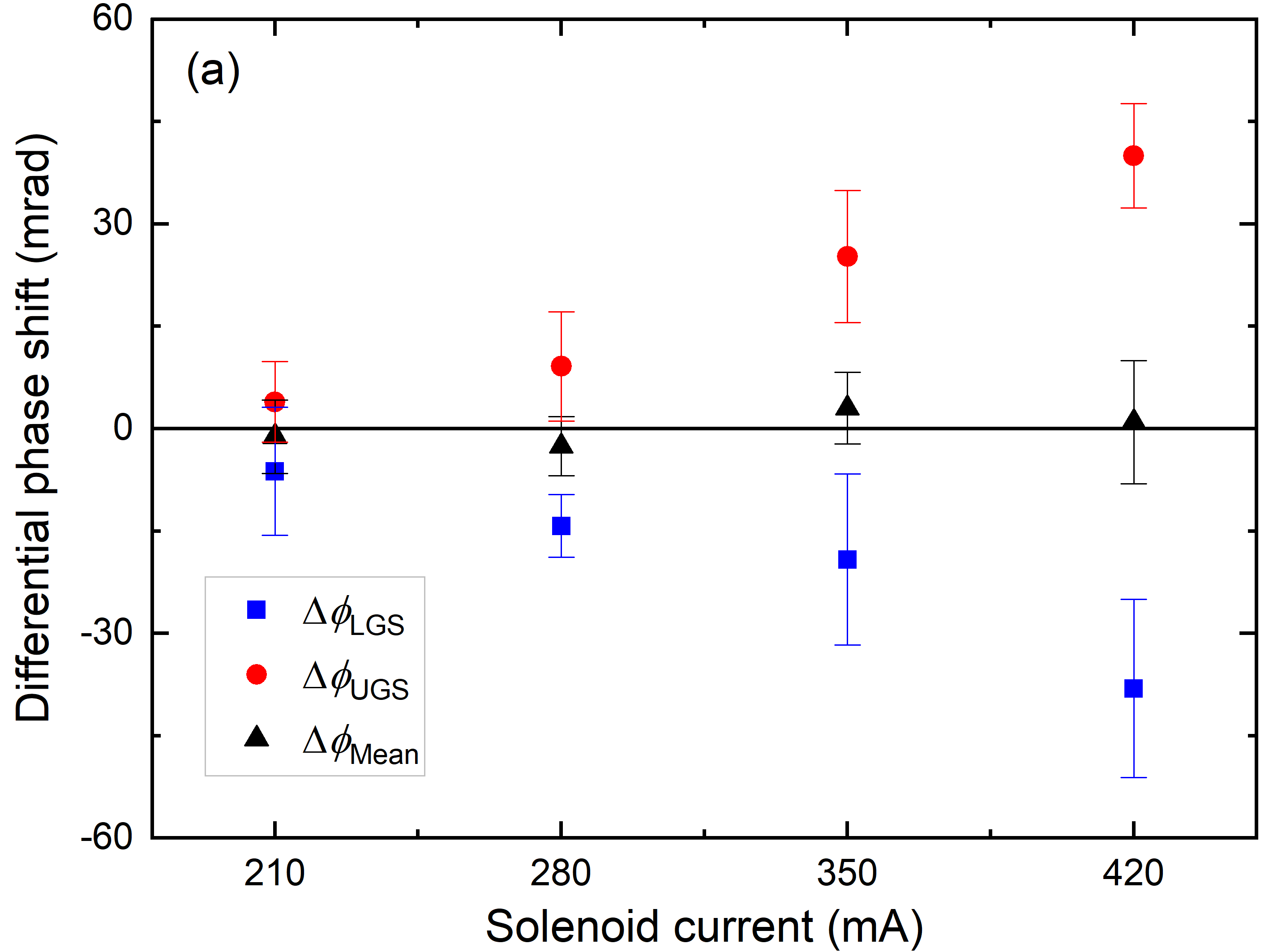}
  \includegraphics[width=0.45\textwidth]{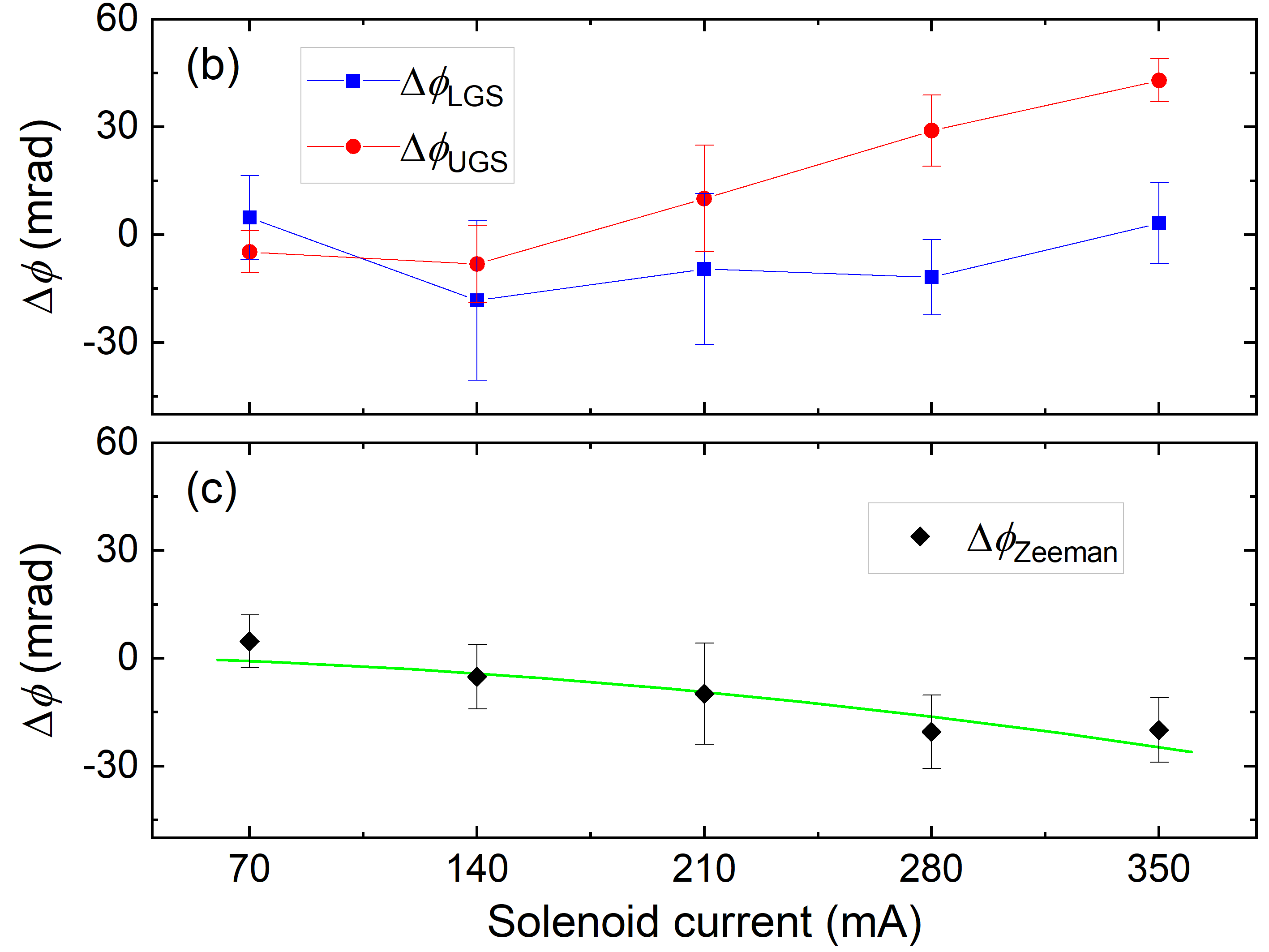}
  \caption{Evaluation based on the HGSE method. (a) The differential phase shifts respond to magnetic field variations simulated by adjusting the solenoid current. The blue, red, and black data are the phases $\Delta \phi_\mathrm{LGS}$, $\Delta \phi_\mathrm{UGS}$, and $\Delta \phi_\mathrm{Mean}=(\Delta \phi_\mathrm{LGS}+\Delta \phi_\mathrm{LGS})/2$, respectively. (b) The phases $\Delta \phi_\mathrm{LGS}$ and $\Delta \phi_\mathrm{UGS}$ contain the modulated gravity-gradient-induced phase shift at each current. (c) The black data are phases $\Delta \phi_{\mathrm{Zeeman}}=(\Delta \phi_\mathrm{LGS}-\Delta \phi_\mathrm{LGS})/2$ extracted from (b). The green curve shows the simulation result without the modulated gravity-gradient-induced phase shift.}
  \label{fig6}
\end{figure*}

At an accuracy of $10^{-11}$ level, all three preceding evaluation methods are effective and accurate. When the accuracy is further improved, the quadratic Zeeman-effect-induced systematic error caused by the variation of the ambient magnetic field will be one of the limitations. The axial shielding factor is less than 10 for our long-baseline magnetic shielding system, which makes it particularly sensitive to the vertical magnetic field. Owing to the presence of metros, elevators, vehicles, and instruments near our laboratory, the maximum variation of the vertical magnetic field is $21\,\mathrm{mG}$. The metros are the most significant factor with the nearest line of $\sim\!200\,\mathrm{m}$ from the laboratory, and their operation causes the magnetic field to fluctuate 12 times more during the day than at night. The laboratory elevator causes a maximum change in the ambient magnetic field of approximately $17\,\mathrm{mG}$ at the top and bottom. However, we always keep it at the bottom of the laboratory while the AI is working to avoid changes in the magnetic field and gradient. Vehicles and instruments cause relatively small variations in ambient magnetic fields, at mG level.

Figure\,\ref{fig5}\textcolor{blue}{(a)} shows the ambient magnetic field in the vertical direction, which is measured by a magnetometer (Bartington Mag690-FL500). During the travelling time of the metros, the maximum variation of the ambient magnetic field is $21\,\mathrm{mG}$ in the vertical direction. We could estimate the residual axial magnetic field inside the shield by the shielding factors, which has a maximum fluctuation of $3.8\,\mathrm{mG}$ in the daytime and $0.2\,\mathrm{mG}$ at night. From Eq.\,(\ref{eq4}), we infer the quadratic Zeeman phase shift based on the magnetic field. Figure\,\ref{fig5}\textcolor{blue}{(b)} shows the corresponding phase shift by the average of the estimated residual magnetic field over 30\,minutes. The shift arises from the average value of the magnetic field and the uncertainty arises from the variation of the magnetic field. Changes in the ambient magnetic field would alter both shift and uncertainty, as shown in Fig.\,\ref{fig5}\textcolor{blue}{(b)}. The ambient magnetic field is stable for approximately 5 hours within a day, and during this period the evaluation method by mapping the absolute magnetic field is accurate, although it takes more time. During the rest of the day, it is difficult to provide an accurate evaluation because of the variation and drift of the magnetic field.

Such interferometers using atoms in the lower and upper hyperfine ground states as co-magnetometers can suppress the effect of strong environmental magnetic fields and field gradients \cite{panda_atomic_2023}. We simulate the fluctuation of the ambient magnetic field by adjusting the solenoid current and demonstrateed that the interferometer with the HGSE method can still measure correctly in a variable ambient magnetic field. As shown in Fig.\,\ref{fig6}\textcolor{blue}{(a)}, taking the average of the two measurements means that the interferometer phases $\Delta \phi_{\mathrm{mean}} = (\Delta \phi_{\mathrm{LGS}} + \Delta \phi_{\mathrm{UGS}})/2$ are stable in a variable magnetic field. The HGSE method can effectively reduce both the systematic shift and uncertainty induced by the quadratic Zeeman effect in a simultaneous $^{85}$Rb-$^{87}$Rb dual-species AI. This method is highly valuable for improving the accuracy of the long-baseline AI, which is easily disturbed by the fluctuation of the ambient magnetic field.

Furthermore, the HGSE method can evaluate systematic errors in real time, avoiding inaccuracies owing to slow temporal changes in ambient magnetic fields and other systematic errors irrelevant to the $F$-state. To further illustrate, we modulate the gravity-gradient-induced phase shift at each current by changing the initial kinematic differences between the $^{85}$Rb and $^{87}$Rb atomic clouds, as shown in Fig.\,\ref{fig6}\textcolor{blue}{(b)}. Nevertheless, the extracted phase shifts $\Delta \phi_{\mathrm{Zeeman}}$ by the HGSE method are consistent with the simulation result without the modulated gravity-gradient-induced phase shift, as shown in Fig.\,\ref{fig6}\textcolor{blue}{(c)}. The HGSE method accurately obtains the evaluation result by subtracting other systematic errors irrelevant to the $F$-state, such as the gravity gradient effect.

\section{CONCLUSION}\label{section6}

In conclusion, we have investigated the Zeeman-effect-induced systematic error in the $^{85}$Rb-$^{87}$Rb dual-species AI. By analyzing the characteristics of the Zeeman-effect-induced phase shift, we realize a HGSE method to evaluate this systematic error. This method obtains evaluation results in real time, avoiding inaccuracies due to slow temporal variation in ambient magnetic fields and other systematic errors irrelevant to the hyperfine ground states. In addition, the HGSE method does not require substantial time to measure the magnetic field and is certainly not related to the accuracy of the magnetic field measurement. The evaluation results of the HGSE method match well with the results of two other methods, which are mapping the magnetic field in the interference region and performing phase measurements at different bias fields. Furthermore, the HGSE method can effectively reduce both systematic shift and uncertainty. The strategy presented in this paper can be used in AIs, especially in long-baseline AIs \cite{zhan_zaiga_2020, badurina_aion_2020, abe_matter-wave_2021, beaufils_cold-atom_2022} for high-precision measurement.

\section*{Acknowledgments}
The authors thank Feng Yang for his useful discussion on the performance of magnetic shielding system. This work was supported by the Hubei Provincial Science and technology major project (Grant No. ZDZX2022000001), the Natural Science Foundation of Hubei Province (Grant No. 2022CFA096), the Innovation Program for Quantum Science and Technology (Grant No. 2021ZD0300603), the Chinese Academy of Sciences Project for Young Scientists in Basic Research (Grant No. YSBR-055), and National Natural Science Foundation of China (Grant No. 12174403, 12304547).

%
\clearpage
\end{document}